# 1. Introduction

Methods of digital topology are widely used in various image processing operations including topology-preserving thinning, skeletonization, simplification, border and surface tracing and region filling and growing.

Usually, transformations of digital objects preserve topological properties. One of the ways to do this is to use simple points, edges and cliques: loosely speaking, a point or an edge of a digital object is called simple if it can be deleted from this object without altering topology. The detection of simple points, edges and cliques is extremely important in image thinning, where a digital image of an object gets reduced to its skeleton with the same topological features.

The notion of a simple point was introduced by Rosenfeld [14]. Since then due to its importance, characterizations of simple points in two, three, and four dimensions and algorithms for their detection have been studied in the framework of digital topology by many researchers (see, e.g., [1, 5, 12-13]). Local characterizations of simple points in three dimensions and efficient detection algorithms are particularly essential in such areas as medical image processing [2, 7-8, 15], where the shape correctness is required on the one hand and the image acquisition process is sensitive to the errors produced by the image noise, geometric distortions in the images, subject motion, etc, on the other hand.

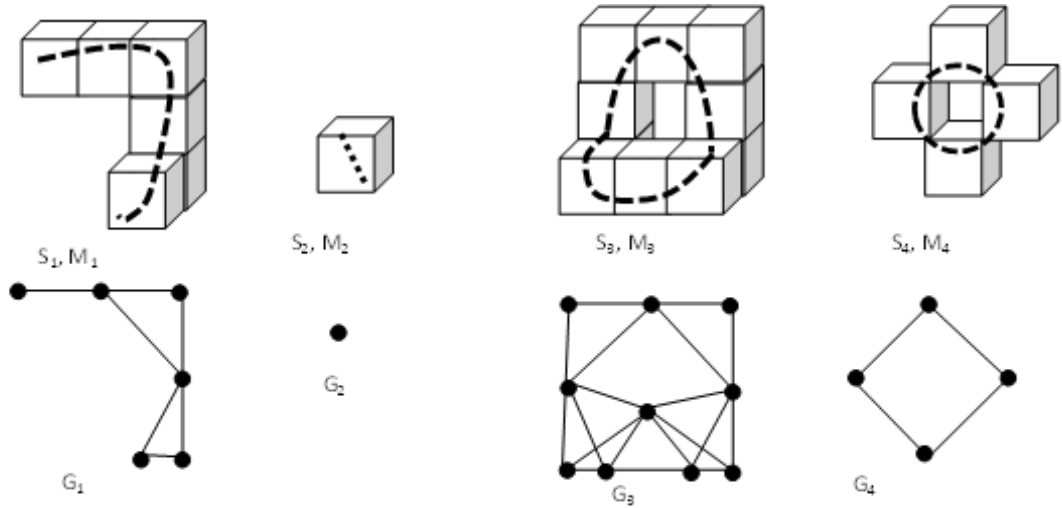

Figure 1. $S_1$ and $S_2$ are segments. $M_1$ and $M_2$ are the cubical models of $S_1$ and $S_2$. $G_1$ and $G_2$ are the intersection graphs of $M_1$ and $M_2$. $G_2$ is homotopy equivalent to $G_1$. $S_3$ and $S_4$ are circles. $M_3$ and $M_4$ are the cubical models of $S_3$ and $S_4$. $G_3$ and $G_4$ are the intersection graphs of $M_3$ and $M_4$.

It has to be noticed that in this paper we use an approach that was developed in [3]. Digital spaces and manifolds are defined axiomatically as specialization graphs with the structure defined by the construction of the nearest neighborhood of every point of the graph. In this approach, the notions of a digital space, a simple point or a simple set are different from those usually to be found in papers on digital topology including papers mentioned above.

This paper presents the notion of a simple pair of points based on digital contractible spaces and contractible transformations of digital spaces. Some new properties of digital n-manifolds which are digital models of continuous n-dimensional manifolds are investigated in section 3. In particular, it is shown that M is a digital n-sphere if for any contractible subspace G, the subspace M-G is contractible.

Section 4 introduces the notions of the simple splitting of a point and the contraction simple pair of points. It is shown that these transformations convert a given digital space to a homotopy equivalent digital space. Based on the simple contraction, we prove that a digital n-sphere S contained in a digital (n+1)-sphere M is a separating space for M. We show that a digital n-manifold M (which does not contain simple points at all) can be transformed to a digital n-manifold N with the minimal number of points (skeleton) by sequential contracting simple pairs.

# 2. Computer experiments as background for digital spaces

The following surprising fact was noticed in computer experiments described in [9]. Suppose that S is a surface in Euclidean space $E^n$. Divide $E^n$ into a set F of cubes with the edge length equal to L and vertex coordinates equal to nL. Call the cubical model of S the family M of cubes intersecting S, and the digital model of S the intersection graph G of M. Suppose that $S_1$ and $S_2$ are isomorphic surfaces, and $G_1$ and $G_2$ are their digital models.

It was revealed that there exists $L_0$ such that for any $L<L_0$, digital models $G_1$ and $G_2$ can be transformed from one to the other with some kind of transformations called contractible.

It is possible to assume that the digital model contains topological and perhaps geometrical characteristics of the surface S. Otherwise, the digital model G is a discrete counterpart of a continuous space S.

To illustrate these experiments, consider examples depicted in fig. 1 and 2. In fig. 1, $S_1$ and $S_2$ are segments,

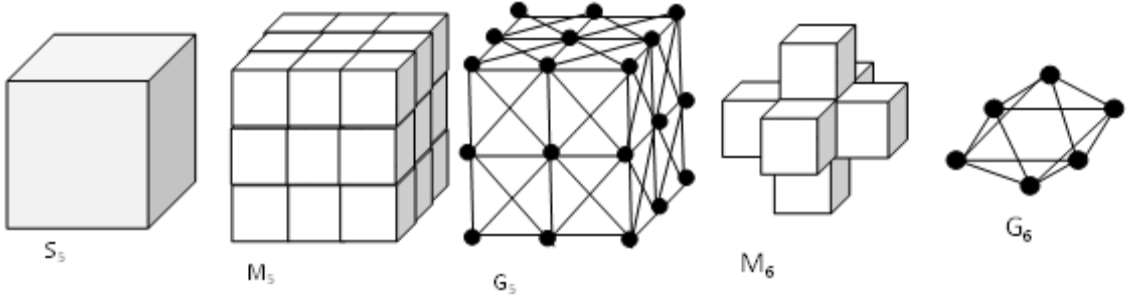

Figure 2. $S_5$ is a topological sphere. $M_5$ and $M_6$ are the cubical models of $S_5$. $G_5$ and $G_6$ are the intersection graphs of $M_5$ and $M_6$.

$M_1$ and $M_2$ are the cubical models of $S_1$ and $S_2$, $G_1$ and $G_2$ are the intersection graphs of $M_1$ and $M_2$. $G_2$ is homotopy equivalent to $G_1$. For circles $S_3$ and $S_4$ shown in fig. 1, $M_3$ and $M_4$ are the cubical models of $S_3$ and $S_4$. $G_3$ and $G_4$ are the intersection graphs of $M_3$ and $M_4$. $G_3$ is homotopy equivalent to $G_4$. $G_3$ is homotopy equivalent to $G_4$, and $G_4$ is a minimal digital 1-dimensional sphere. A topological sphere $S_5$ in fig.2 is the surface of some cube. $M_5$ and $M_6$ are the cubical models of $S_{55}$, $G_5$ and $G_6$ are the intersection graphs of $M_5$ and $M_6$. $G_5$ is homotopy equivalent to $G_6$ which is a minimal digital 2-dimensional sphere.

**3. Contractible graphs and contractible transformations. Digital n-manifolds.**

In order to make this paper self-contained we will summarize the necessary information from previous papers.

By a graph we mean a simple undirected graph G=(V,W), where $V=\{v_1,v_2,...v_n,…\}$ is a finite or countable set of points, and $W = \{(v_pv_q),....\}\subseteq V\times V$ is a set of edges. Such notions as the connectedness, the adjacency, the dimensionality and the distance on a graph G are completely defined by sets V and W. We use the notations $v_p\in G$ and $(v_pv_q)\in G$ if $v_p\in V$ and $(v_pv_q)\in W$ respectively if no confusion can result. |G| denotes the number of points in G.

Since in this paper we use only subgraphs induced by a set of points, we use the word subgraph for an induced subgraph. We write H⊆G. Let G be a graph and H⊆G. G-H will denote a subgraph of G obtained from G by deleting all points belonging to H. For two graphs G=(X,U) and H=(Y,W) with disjoint point sets X and Y, their join G⊕H is the graph that contains G, H and edges joining every point in G with every point in H. The subgraph O(v)⊆G containing all points adjacent to v (without v) is called the rim or the neighborhood of point v in G, the subgraph U(v)=v⊕O(v) is called the ball of v. Graphs can be transformed from one into another in a variety of ways. Contractible transformations of graphs seem to play the same role in this approach as a homotopy in algebraic topology [10-11].

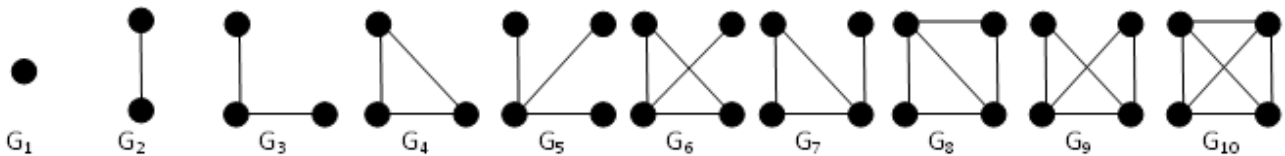

Figure 3. Contractible graphs with the number of points n<5.

**Definition 3.1.**
- A graph G is called contractible (fig. 3), if it can be converted to the trivial graph K(1) by sequential deleting simple points.
- A point v of a graph G is said to be simple if its rim O(v) is a contractible graph.

An edge (vu) of a graph G is said to be simple if the joint rim O(vu)=O(v)∩O(u) is a contractible graph. In [10], it was shown that if (vu) is a simple edge of a contractible graph G, then G-(vu) is a contractible graph. Thus, a contractible graph can be converted to a point by sequential deleting simple points and edges. In fig.3, $G_{10}$ can be converted to $G_9$ or $G_8$ by deleting a simple edge. $G_9$ can be converted to $G_7$ or $G_6$ by deleting a simple edge. $G_6$ can be converted to $G_5$ by deleting a simple edge. $G_7$ can be converted to $G_4$ by deleting a simple point. $G_5$ can be converted to $G_3$ by deleting a simple point. $G_3$ can be converted to $G_2$ by deleting a simple point. $G_2$ can be converted to $G_1$ by deleting a simple point.

Deletions and attachments of simple points and edges are called contractible transformations. Graphs G and

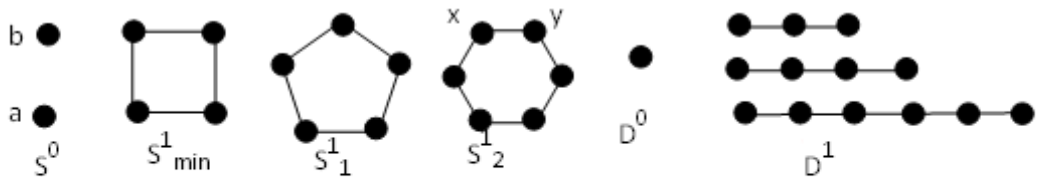

Figure 4. Zero- and one-dimensional spheres $S^0$ and $S^1$ and zero- and one-dimensional disks $D^0$ and $D^1$. {x,y} is a simple pair.

H are called homotopy equivalent or homotopic if one of them can be converted to the other one by a sequence of contractible transformations.
Homotopy is an equivalence relation among graphs. Contractible transformations retain the Euler characteristic and homology groups of a graph [9].
Properties of graphs that we will need in this paper were studied in [9-10].

**Proposition 3.1.**
- Let G be a graph and v be a point (v∉G). Then the cone v⊕G is a contractible graph.
- Let G be a contractible graph and S(a,b) be a disconnected graph with just two points a and b. Then S(a,b)⊕G is a contractible graph.
- Let G be a contractible graph with the cardinality |G|>1. Then it has at least two simple points.
- Let H be a contractible subgraph of a contractible graph G. Then G can be transformed into H by sequential deleting simple points.
- Let graphs G and H be homotopy equivalent. G is connected if and only if H is connected. Any contractible graph is connected.

Further on, if we consider a graph together with the natural topology on it, we will use the phrase 'digital space". We say "space" to abbreviate "digital space", if no confusion can result. Let us recall some useful properties of digital n-spheres and n-manifolds.

**Definition 3.2.**
- A *digital 0-dimensional sphere* is a disconnected digital space $S^0(x,y)$ with just two points x and y.
- A connected space M is called a *digital n-sphere, n>0,* if for any point v∈M, the rim O(v) is a digital (n-1)-sphere, and for some point v, the space M-v is contractible (see [3-47]).
- The join $S^n_{min}=S^0_1 \oplus S^0_2 \oplus \ldots S^0_{n+1}$ of (n+1) copies of the zero-dimensional surface $S^0$ is called a minimal n-sphere.

A digital 0-sphere $S^0$ and digital 1-spheres $S^1_{min}$, $S^1_1$ and $S^1_2$ are depicted in fig. 4. Figure 5 shows digital 2-spheres $S^2_{min}$, $S^2_1$ and $S^2_2$. Graphs that model digital minimal 1-, 2- and 3-dimensional spheres $S^1_{min}$, $S^2_{min}$ and $S^3_{min}$ are shown in fig. 6.

**Proposition 3.2 ([6]).**
   Let M be a digital n-sphere, n>1. Then:
   (a) For any point v∈M, the space M-v is contractible.
   (b) For any contractible space G⊆M, the space M-G is contractible
   (c) The join $S^0(x,y)\oplus M$ is a digital (n+1)-sphere.

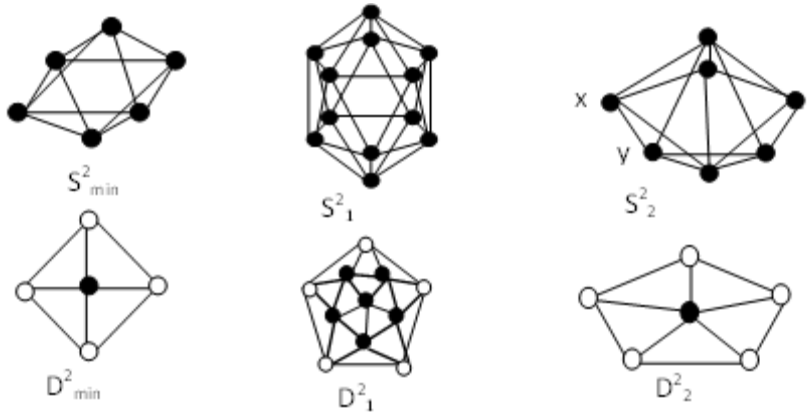

Figure 5. Digital 2-spheres and 2-disks. {x,y} is a simple pair.

   (d) M is homotopy equivalent to $S^n_{min}$.
To make the reading easier, we have presented the proof in Appendix 1.

**Definition 3.3.**
   Let M be a digital n-sphere, and v be a point of M. A contractible space D=∂D∪IntD=M-v is called a digital n-disk *with the boundary* ∂N=O(v) and the interior IntD=M-U(v).

A digital 0-disk $D^0$ and digital 1-disks $D^1$ are depicted in fig. 4. Figure 5 shows digital 2-disks $D^2_{min}$, $D^2_1$ and $D^2_2$.

The following property is a consequence of definition 3.3

**Corollary 3.1.**
   Let D=∂D∪IntD be a digital n-disk. If a point x∈IntD, then the rim O(x) is a digital (n-1)-sphere, if a point x∈∂D, the rim O(x) is a digital (n-1)-disk.

**Definition 3.4.**
A connected space M is called a digital n-dimensional manifold, n>1, if the rim O(v) of any point v is a digital (n-1)-dimensional sphere.

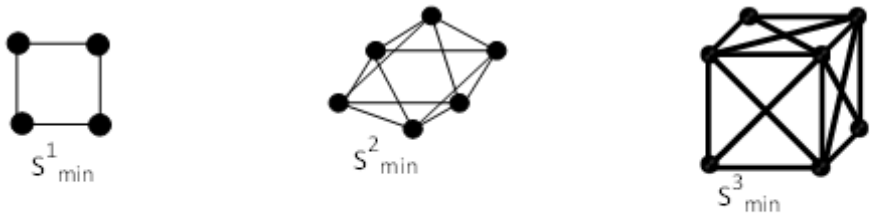

Figure 6. Minimal 1-, 2- and 3-dimensional spheres.

A digital n-sphere is a digital n-manifold. Digital 2-manifolds: a torus T and a projective plane P are depicted in fig. 7. Notice that T contains sixteen points, P contains eleven points.
Consider difference between a digital n-sphere and a digital n-manifold which is not a sphere.

**Proposition 3.3 ([6]).**
>   Let M be a digital n-manifold, G be a contractible subspace of M and v be a point in M. Then subspaces M-G and M-v are homotopy equivalent to each other.

The proof is to be found in Appendix 1.

**Corollary 3.3.**
>   Let M be a digital n-manifold and G be a contractible subspace of M. M is a digital n-sphere if and only if the space M-G is contractible.

Figure 7 illustrates proposition 3.3 and corollary 3.3. Since T is not a digital 2-sphere then T without a point {7} is not a contractible space. It is easy to check directly that T-{7} shown in fig. 7 is homotopy equivalent to the space E, i.e, T-{7} can be converted to E by sequential deleting simple points and edges. Similarly, since a projective plane P is not a digital 2-sphere then P without a point {a} is not a contractible space. One can check directly that P-{a} shown in fig. 7 is homotopy equivalent to the space C which is a minimal digital 1-sphere. One can see that there is a full correspondence between digital topology results for a torus and a projective plane and classical topology results.

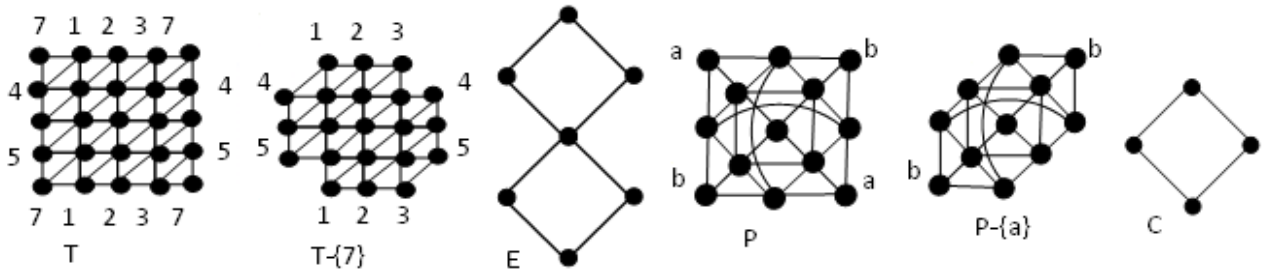

Figure 7. A digital 2-dimensional torus T and a digital 2-dimensional projective plane P. T-{7} and P-{a} are not contractible spaces. By sequential deleting simple points and edges, T-{7} can be converted to E, and P-{a} can be converted to C.

**4. Simple pairs of points of graphs and digital n-manifolds**

In graph theory, the contraction of points x and y in a graph G is the replacement of x and y with a point z such that z is adjacent to the points to which points x and y were adjacent. In paper [6], the contraction of simple pairs of points was used for classification of digital n-manifolds.

**Definition 4.1.**
- Let G be a graph and x and y be adjacent points of G. We say that {x,y} is a simple pair if any point v belonging to U(x)-U(y) is not adjacent to any point u belonging to U(y)-U(x).
- Let G be a graph and {x,y} be a simple pair of G. The replacement of x and y with a point z such that O(z)=U(x)∪U(y)-{x,y} is called the simple contraction of points x and y or F-transformation. FG=(G∪z)-{x,y} is the graph that results from contracting points x and y.
- Let G be a graph and z be a point of G. The replacement of z with adjacent points x and y in such a way that U(x)∪U(y)-{x,y}=O(z), and any point v belonging to U(x)-U(y) is not adjacent to any point u belonging to U(y)-U(x) is called the simple splitting of z or R-transformation. RG=(G∪{x,y})-z is the graph that results from simple splitting point z.

Simple F- and R-transformations are invertible. For a given F-transformation, the inverse of F is a simple splitting R=F$^{-1}$. In fig. 8(a), {x,y} is a simple pair of points lying in some graph G. Fig. 8(b) shows a part of H=FG=(G∪z)-{x,y} obtained by {x,y} contraction. A pair {a,b} depicted in fig. 8(c) is not a simple pair. It has to be noticed that this definition of a simple pair is different from the one proposed in [20]. The following corollary is an obvious consequence of definition 4.1 (see fig. 8). In fact, fig. 8 illustrates the proof of corollary 4.1.

**Corollary 4.1.**
Let G be a graph, and adjacent points x and y belong to G. {x,y} is a simple pair of points if and only if there is no digital minimal 1-sphere S$^1$={x,y,a,b} lying in G and containing points x and y.

**Proposition 4.1.**
Let {x,y} be a simple pair lying in a graph G. Then the graph H=(G-{x,y})∪z obtained by the contraction of {x,y} is homotopy equivalent to G.
**Proof.**
First, show that the graph B=U(x)∪U(y) is contractible. Pick a point v∈U(x)-U(y). Since v is not adjacent to any point u belonging to U(y)-U(x) then the rim O$_B$(v) of v is the cone x⊕(O(xv), i.e., a contractible graph. Therefore, v is a simple point of B, and can be deleted from B. For the same reason, all points belonging to U(x)-U(y) can be deleted from B by sequential deleting simple point. The obtained graph U(y)=y⊕(O(y) is homotopy equivalent to B. Since U(y) is a contractible graph according to proposition 2.1, then B=U(x)∪U(y) is a contractible graph.
Glue a simple point z to G in such a way that O(z)=U(x)∪U(y). In the obtained graph P=G∪z, the rim of x is the cone O$_P$(x)=z⊕O(x). Therefore, point x is simple in P and can be deleted from P. In the obtained graph Q=P-{x}, the rim of y is the cone O$_Q$(y)=z⊕(O(y)-{x}). Therefore, y is simple in Q and can be deleted from Q. The obtained graph Q-{y}=H=(G-{x,y})∪z is homotopy equivalent to G. The proof is complete.□

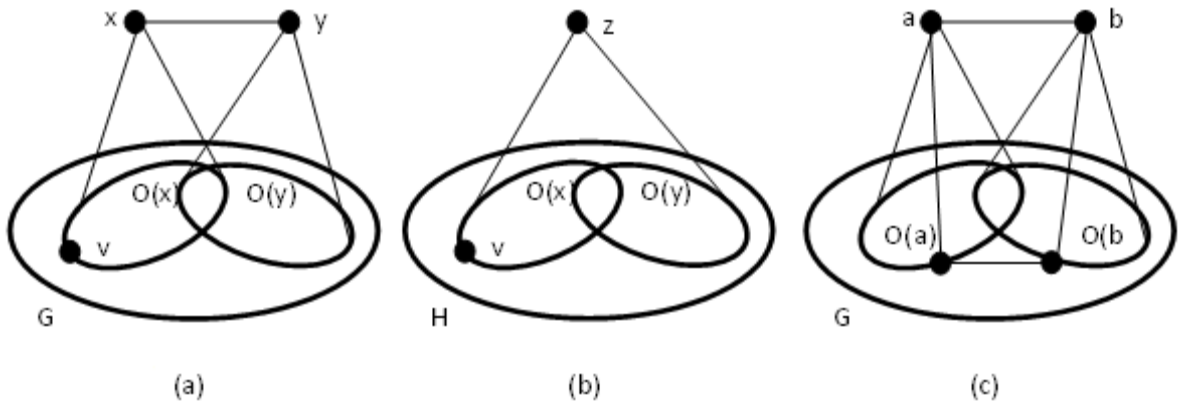

Figure 8. (a) {x,y} is a simple pair. (b) O(z)=O(x)∪O(y)-{x,y}. (c) {a,b} is not a simple pair.

It follows from proposition 4.1 that a contractible graph can be converted to a one-point graph by a sequence of simple contraction (see fig. 3).

**Proposition 4.2.**
Let {x,y} be a simple pair of a graph G, and the graph H=(G-{x,y})∪z be obtained by the contraction of {x,y}. Then:
  (a) If P is a subgraph of G containing {x,y} then {x,y} is a simple pair of P, and the graph Q=(P-{x,y})∪z is homotopy equivalent to P.
  (b) If v is a point of G-{x,y}, and v∉O(x)∩O(y) then the rim O$_M$(v) in M is isomorphic to the rim O$_N$(v) in N.
**Proof.**

(a) Evidently, $U_H(x)=U(x)\cap H$, $U_H(y)=U(y)\cap H$. Since any point v belonging to $U(x)-U(y)$ is not adjacent to any point u belonging to $U(y)-U(x)$, then any point v belonging to $(U(x)-U(y))\cap H$ is not adjacent to any point u belonging to $(U(y)-U(x))\cap H$. Therefore, $\{x,y\}$ is a simple pair of H.

(b) To prove (b), consider first a point v belonging to $O(x)$ (see fig. 8(a-b)). It follows directly from the structure of N that the rim $O_N(v)$ is obtained from $O_M(v)$ by replacing x with z. Therefore, $O_N(v)$ is isomorphic to $O_M(v)$. If $v \in M-(U(x)\cup U(y))$ then $O_N(v)=O_M(v)$. The proof is complete. ☐

The advantage of using contractions of simple pairs of points is that they retain global as wel as local topology of digital n-manifolds. They necessarily convert a digital n-manifld M to a digital n-manifold N which is homotopy equivalent to M. Consider properties of simple pairs lying in a digital n-sphere or a digital n-manifold M. Proposition 4.3 directly follows from corollary 4.1.

**Proposition 4.3.**
> Let M be a digital n-manifold, and adjacent points x and y belong to M. $\{x,y\}$ is a simple pair of points if and only if there is no digital minimal 1-sphere $S^1=\{x,y,a,b\}$ lying in M and containing points x and y.

Notice that a digital n-manifold has no simple points and simple edges. Nevertheless, its number of points can be reduced by the contraction of a simple pair of points.

**Proposition 4.4.**
> (a) A minimal digital n-sphere $S^n_{min}=S^0_1 \oplus S^0_2 \oplus \ldots S^0_{n+1}$ has no simple pairs of points.
> (b) Let M be a digital n-sphere, n>0, and $\{x,y\}$ be a simple pair lying in M. Then $U(x)\cup U(y)=D=\partial D\cup IntD$ is a digital n-disk with the boundary $\partial D=U(x)\cup U(y)-\{x,y\}$ and the interior $IntD=\{x,y\}$.
> (c) Let M be a digital n-sphere, n>0, $\{x,y\}$ be a simple pair lying in M, and $N=FM=(M\cup z)-\{x,y\}$ be the space obtained by the contraction of $\{x,y\}$. Then $N=(M\cup z)-\{x,y\}$ is a digital n-sphere.
> (d) Let $D=\partial D\cup IntD$ be a digital n-disk. If $|IntD|>1$, then IntD contains a simple pair.

**Proof.**
(a) Assertion (a) follows from construction of $S^n_{min}$ (see figure 6).
(b) By construction of $U(x)\cup U(y)$, $U(x)\cup U(y)$ is a contractible space, the rims of points x and y in $U(x)\cup U(y)$ are digital (n-1)-spheres, the rim of any point v belonging to $U(x)\cup U(y)-\{x,y\}$ is a digital (n-1)-disk. It follows from the structure of $U(x)\cup U(y)-\{x,y\}$ that $U(x)\cup U(y)-\{x,y\}$ is a digtal (n-1)-sphere. Thus, $U(x)\cup U(y)$ is a digital n-disk according to definition 3.3.
(c) The proof is by induction on the dimension n. For n=1, the assertion is verified directly as one can see in fig. 4, where $\{x,y\}\subseteq S^1_2$ and $S^1_1=(S^1_2\cup z)-\{x,y\}$. Assume that the assertion is valid whenever n<k. Let n=k. Show first that the rim of any point of N is a digital (n-1)-sphere. For $z\in N$, $O(z)=U(x)\cup U(y)-\{x,y\}=(O(x)-y)\#(O(y)-x)$ is a digital (n-1)-sphere according to assertion (b). For $v\in(O(x)-O(y))\subseteq N$, $O(v)$ in N is isomorphic to $O(v)$ in M, i.e., a digital (n-1)-sphere. For $v\in O(x)\cap O(y)\subseteq N$, $O(v)$ in M is a digital (n-1) sphere containing a simple pair $\{x,y\}$. Therefore, $O(v)$ in N is a digital (n-1) sphere by the induction hypothesis. For $v\in N-(U(x)\cup U(y))$, $O(v)$ in N is the same as $O(v)$ in M, i.e., a digital (n-1)-sphere. To show that N-u is a contractible space, pick a point $u\in N-(U(x)\cup U(y))$. M-u is a contractible space according to theorem 3.1. N-u=F(M-u) is homotopy equivalent to M-u according to proposition 3.2. Hence, N-u is a contractible space. Thus, N is a digital n-sphere.
(d) The proof of assertion (d) is similar to the proof of (c) and is omitted. ☐

In fig. 4, a digital 1-sphere $S^1_2$ contains a simple pair $\{x,y\}$. Evidently, $(S^1_2\cup z)-\{x,y\}$ is $S^1_1$. A digital 2-sphere $S^2_2$ depicted in fig. 5 contains a simple pair $\{x,y\}$. Contracting $\{x,y\}$ converts $S^2_2$ to $S^2_{min}$. All minimal n-spheres depicted in fig. 6 do not contain simple pairs.

**Definition 4.4.**
> Let A and B be subspaces of a connected space M. A and B are called separated if any point in A is non-adjacent to any point in B.

If a connected space M is represented as the union A∪C∪B, where spaces A and B are separated, we will say that the union M=A∪C∪B is a separation of M by the space C and C is a separating space for M.

**Proposition 4.5.**
Let M be a digital n-sphere and S be a digital (n-1)-sphere in M, S⊆M. Then M=G∪S∪H is the separation of M by S and G∪S and S∪H are n-disks.

Proof.
The proof is by induction on the number of points |M|=k of M. For k=2n+2, M is a minimal digital n-sphere $M=S^0_1 \oplus S^0_2 \oplus \ldots S^0_{n+1} = S^0(v,u) \oplus S^0_2 \oplus \ldots S^0_{n+1} = S^0(v,u) \oplus S^{n-1}_{min} = v \cup S^{n-1}_{min} \cup u$. Assume that the proposition is valid whenever k<s. Let k=s.

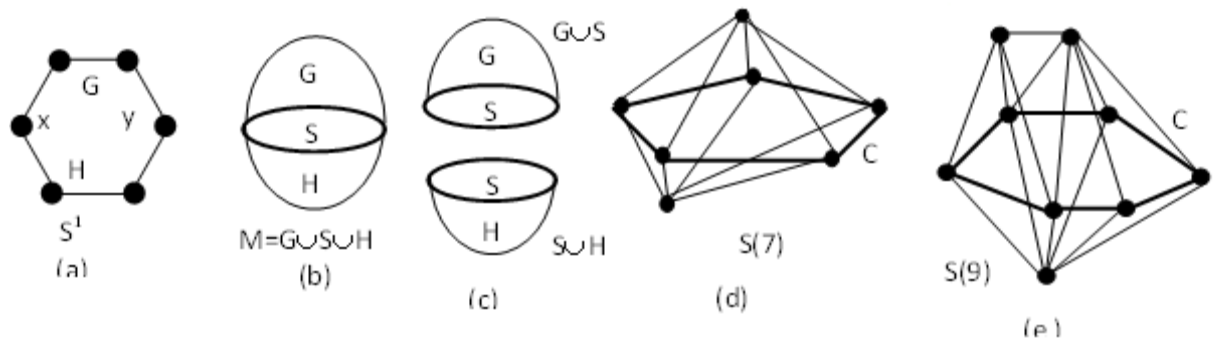

Figure 9. (a) A digital 0-sphere $S^0=\{x,y\}$ is a separating space in a 1-sphere $S^1$. (b) The separation of M by S. (c) G∪S and S∪H are digital n-disks. (d)- (e) A digital 1-sphere C is a separating space in 2-spheres S(7) and S(9).

Suppose that a simple pair {x,y}⊆S. Then N=FM=(M∪{z})-{x,y} is a digital n-sphere, $S_1$=FS=(S∪{z})-{x,y} is a digital (n-1)-sphere, and $S_1$⊆N. Therefore, $S_1$ is a separating space for N=$G_1$∪$S_1$∪$H_1$ by the induction hypothesis, and $G_1$∪$S_1$ and $S_1$∪$H_1$ are digital n-disks. According to propositions 3.1 and 3.2, $F^{-1}(G_1 \cup S_1)$=G∪S and $F^{-1}(S_1 \cup H_1)$= S∪H are digital n-disks, and S is a separating space for M=$F^{-1}$N. Suppose that a simple pair {x,y}⊆M-S. The proof is very similar to the above proof and is omitted. Suppose now that a simple pair x∈S, y∈M-S. The proof is also very similar to the above proof and is omitted. The proof is complete. ☐

In fig. 9(b-c), S is a separating space for a digital n-sphere M. G∪S and S∪H are digital n-disks. A digital 0-sphere $S^0(x,y)$ separates a digital 1-sphere $S^1$=G∪$S^0(x,y)$∪H shown in fig. 8(a). A digital 1-sphere C is a separating space in 2-spheres S(7) and S(9) depicted in fig. 8(d-e). The following corollary is a consequence of proposition 4.5.

**Corollary 4.2.**
Let D=∂D∪IntD and E=∂E∪IntE be digital n-disks. If ∂D and ∂E are isomorphic, f: ∂D→ ∂E, then the space D#E obtained by identifying points belonging to ∂D with corresponding points belonging to ∂E is a digital n-sphere.

Figure 9(c) shows two n-disks G∪S and S∪H. Their connected sum (G∪S)#( S∪H) is a digital n-sphere M shown in fig. 9(b).

**Definition 4.6.**
A digital n-manifold is called compressed if it does not contain simple pairs of points.

It is clear that a digital n-sphere is a compressed digital n-manifold as one can see in fig. 6. It is easy to check directly that a digital 2-torus T and a digital projective plane P depicted in fig. 7 are compressed digital 2-manifolds.

The following assertion is obvious.

**Proposition 4.5.**
A digital manifold M can be converted to a compressed form CM by sequential contracting simple pairs of points.

If CM is a compressed digital n-manifold obtained from M by sequential contracting simple pairs of points then |CM|≤|M|. Although CM has the same topology as M, in some sense, CM is "simpler" then M. Therefore, CM can be considered as the representative of the class all digital n-manifolds which are homotopy equivalent to CM.

**Conclusion.**
Sometimes, it is not sufficient to use deletions of simple points, edges and cliques for topology-preserving thinning digital objects because digital objects have no simple points and edges at all. Contractions of simple pairs of points enable to reduce the number of points of a digital space to the minimum while preserving local and global topology.

**References.**

[1]   Y. Bai, X, Han, J. L. Prince, Digital topology on adaptive octree grids, Journal of Mathematical Imaging and Vision 34 (2) (2009) 165-184.
[2]   P. Bazin and D.L.Pham, Topology-preserving tissue classification of magnetic resonance brain images, IEEE Transactions on Medical Imaging 26 (4) (2007) 487-496.
[3]   A.V. Evako, R. Kopperman, Y.V. Mukhin, "Dimensional properties of graphs and digital spaces, Journal of Mathematical Imaging and Vision 6 (1996) 109-119.
[4]   A. V. Evako, Topological properties of closed digital spaces. One method of constructing digital models of closed continuous spaces by using covers, Computer Vision and Image Understanding 102 (2006) 134-144.
[5]   A. V. Evako, Characterizations of simple points, simple edges and simple cliques of digital spaces. One method of topology-preserving transformations of digital spaces by deleting simple points and edges, Graphical Models. 73 (2011) 1-9.
[6]   A. V. Evako, Classification of digital n-manifolds, Discrete Applied Mathematics, In press, DOI: 10.1016/j.dam.2014.08.023.
[7]   X. Han, C. Xu, U. Braga-Neto, and J. L. Prince, Topology correction in brain cortex segmentation using a multiscale graph-based approach, IEEE Transactions on Medical Imaging 21 (2) (2002) 109-121.
[8]   X. Han, C. Xu, and J. L. Prince, A topology preserving level set method for geometric deformable models, IEEE Transactions on Pattern Analysis and Machine Intelligence 25 (2003) 755–768.
[9]   A.V. Ivashchenko, "Representation of smooth surfaces by graphs. Transformations of graphs which do not change the Euler characteristic of graphs, Discrete Mathematics 122 (1993) 219-233.
[10]  A. V. Ivashchenko, Contractible transformations do not change the homology groups of graphs, Discrete Mathematics 126 (1994) 159-170.
[11]  A. V. Ivashchenko, Some properties of contractible transformations on graphs, Discrete Mathematics 133 (1994) 139-145.
[12]  T. Y. Kong, P. K. Saha, A. Rosenfeld, Strongly normal sets of contractible tiles in N dimensions, Pattern Recognition 40 (2) (2007) 530-543.
[13]  N. Passat, M. Couprie, G. Bertrand, Minimal simple pairs in the 3-d cubic grid, Journal of Mathematical Imaging and Vision 32 (2) (2008) 239 – 249.

[14] A. Rosenfeld, Connectivity in digital pictures, Journal of the Association for Computing Machinery 17 (1970) 146–160.

[15] F. Segonne, J.-P. Pons, E. Grimson, and B. Fischl, Active contours under topology control genus preserving level sets, In Springer, editor, Computer Vision for Biomedical Image Applications 3765 of Lecture Notes in Computer Science (2005) 135-145.